\documentclass{article}

\usepackage{./shared/arxiv}
\usepackage[natbib=true,style=authoryear,doi=true,isbn=true,url=false,eprint=true]{biblatex}
\usepackage[utf8]{inputenc} % allow utf-8 input
\usepackage[T1]{fontenc}    % use 8-bit T1 fonts
\usepackage{hyperref}       % hyperlinks
\usepackage{url}            % simple URL typesetting
\usepackage{booktabs}       % professional-quality tables
\usepackage{amsfonts}       % blackboard math symbols
\usepackage{nicefrac}       % compact symbols for 1/2, etc.
\usepackage{microtype}      % microtypography
\usepackage{graphicx}

\usepackage{doi}

\usepackage{gensymb} % for degree symbol
\usepackage{rotating} % for sideways figures
\usepackage{caption}
\usepackage{subcaption}
\usepackage[nobottomtitles*]{titlesec}
\usepackage{amsmath}
\usepackage{setspace}

\newcommand{\code}[1]{\lstinline[language=bash, basicstyle=\ttfamily\small]|#1|}
\usepackage{color}
\definecolor{lightgrey}{rgb}{0.975, 0.975, 0.975}
\definecolor{midgrey}{rgb}{0.6, 0.6, 0.6}
\definecolor{deepgrey}{rgb}{0.2, 0.2, 0.2}
\definecolor{codegreen}{rgb}{0, 0.7, 0}
\definecolor{codepink}{rgb}{0.8196, 0.10, 0.654}
\usepackage{listings}
\title{Untangling urban data signatures: unsupervised machine learning methods for the detection of urban archetypes at the pedestrian scale}

\date{} % remove the date

\author{
	\href{https://orcid.org/0000-0003-3790-0638}{
		\includegraphics[width=0.25cm, height=0.25cm, keepaspectratio]{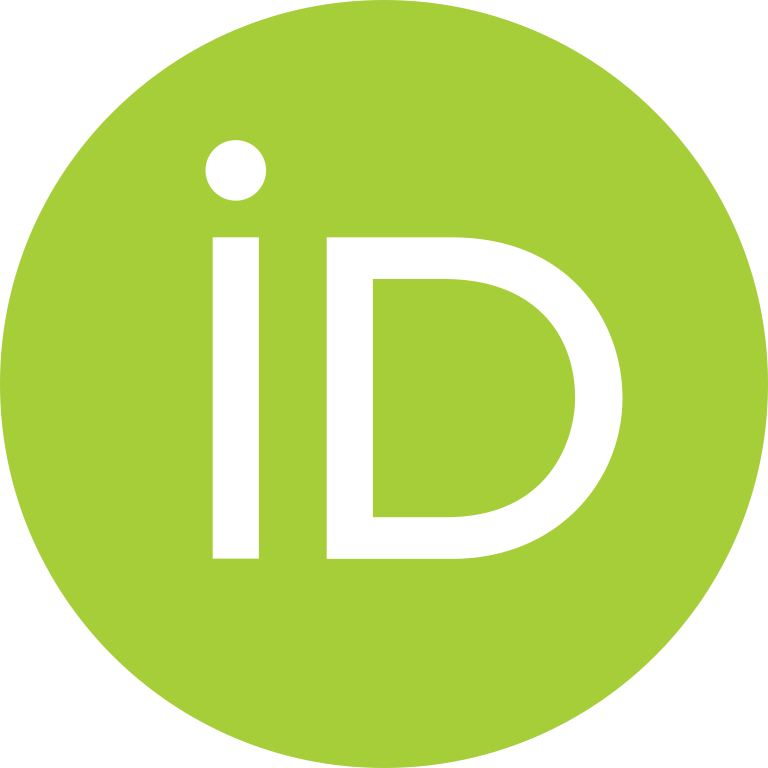}
		\hspace{1mm}Gareth D. Simons
	}
	\thanks{Benchmark Urbanism \texttt{gareth@benchmarkurbanism.com}}
}

\renewcommand{\shorttitle}{Untangling urban data signatures}

\hypersetup{
	pdftitle={
		Untangling urban data signatures: unsupervised machine learning methods for the detection of urban archetypes at the pedestrian scale
	},
	pdfsubject={
		physics.soc-ph
	},
	pdfauthor={
		Gareth~D.~Simons,
	},
	pdfkeywords={
		computation,
		data science,
		land-use analysis,
		machine learning,
		morphometrics,
		network analysis,
		spatial analysis,
		unsupervised ML,
		urban analytics,
		urban planning,
		urban morphology,
		urbanism
	},
}

\begin{document}
\maketitle
\begin{abstract}
	Urban morphological measures applied at a high-resolution of spatial analysis can yield a wealth of data describing characteristics of the urban environment in a substantial degree of detail; however, such forms of high-dimensional numeric datasets are not immediately relatable to broader constructs rooted in conventional conceptions of urbanism. Data science and machine learning (ML) methods provide an opportunity to explore such forms of complex datasets by applying unsupervised ML methods to reduce the dimensionality of the data while recovering latent themes and characteristic patterns which may resonate with urbanist discourse more generally.

Dimensionality reduction and clustering methods, including Principal Component Analysis (PCA), Variational Autoencoders, and an Autoencoder based Gaussian Mixture Model, are discussed and demonstrated for purposes of `untangling' urban datasets, revealing themes bridging quantitative and qualitative descriptions of urbanism. The methods are applied to a dataset for Greater London consisting of network centralities, land-use accessibilities, mixed-use measures, and density measures. The measures are computed at pedestrian walking tolerances at a $20m$ network resolution using the \code{cityseer-api} \code{Python} package, which utilises a local windowing-methodology with distances computed directly over the network and with aggregations performed dynamically and with respect to the direction of approach, thus preserving the relationships between the variables and retaining contextual precision. 

Whereas the demonstrated methods hold tremendous potential, their power is difficult to convey or fully exploit using conventional lower-dimensional visualisation methods, thus underscoring a need for subsequent research into how such methods may be coupled to interactive visualisation tools to further elucidate the richness of the data and its potential implications.
\end{abstract}
\section{Detection of urban archetypes with unsupervised machine learning methods}\label{detection-of-archetypes}

Vibrant pedestrian districts manifest an affinity for complexity and its requisite diversity, as do complex systems more generally \citep{Page2011}. However, urban master plans have historically demonstrated a proclivity towards reductionism. Cities were increasingly rearranged around motor vehicles and reconceived in the abstract on drawings boards; the more granular, dense, mixed-use, and visually `messy' artefacts of evolved cities were swept-aside for grandiose compositions that were ultimately too large, too homogenous, and too resistant towards change for pedestrian-based forms of urbanism to thrive \citep{Harvey1989, Lyon1999}. Though initially associated with high-modernism, aspects of these patterns remain prevalent in forms of contemporary urbanism manifesting across the spectrum from suburbia to romanticised smart city master-plans, explicitly or implicitly emphasising idealised efficiencies at the expense of spatial complexity \citep{Greenfield2013, Townsend2013, Sterling2014}.

The complex systems interpretation of cities, replete with dynamics from emergence to non-linearities to phase-changes \citep{Batty1994, Batty2005, Batty2013, Allen2012}, resists simple averages and crude models \citep{Jacobs1961}. A dilemma therefore faces architects, urban designers, and planners tasked with the challenge of how to plan for inherently unpredictable processes at the urban scale \citep{Portugali2012, Marshall2009}. Whereas it is not possible to anticipate every last action of city citizens --- and how these chains of interaction might coalesce or bifurcate through space --- it is possible to gauge, more generally, how that certain forms of urbanism may be more conducive to large numbers of permutations of complex interactions \citep{Alexander1967}. Complex systems derived methods, including network centralities and mixed-use measures, thus serve as proxies for urban complexity, foreshadowing networks of potential interactions available to city citizens. The question then arises: if we apply pedestrian-scale centrality and mixed-use measures using sufficiently precise and high-resolution analysis, then can emerging forms of data analysis and unsupervised machine learning methods be used to `untangle' and `sift-out' signature patterns from the mass of ensuing data? These terms are used in the literal sense because large and high-dimensional datasets require teasing apart to reveal latent themes, which may ultimately help bridge the gap from quantitative forms of urban analytics to qualitatively framed conceptions of cities \citep{Portugali2012}.

At first glance, the use of data science or machine learning models to understand or even predict aspects of healthy cities may appear ironic and misguided given planning's problematic past. Examples such as Robert Moses' ruthless dismemberment of New York City \citep{Flint2011} and failed housing schemes such Pruitt-Igoe are stark reminders that reliance on shallow or overly abstract interpretations of urbanism can lead to misguided decisions and problematic forms of urban policy. It is precisely these issues that provoked Jane Jacobs' \emph{The Death and Life of Great American Cities} \citep{Jacobs1961} which would forever change perspectives on urbanism. Jacobs laments that planners had misconstrued the nature of cities and had mistakenly assumed that decisions made in the abstract were somehow sufficient to deal with the emergent complexity of healthy cities. She articulates her thoughts with reference to Warren Weaver's seminal paper \emph{Science and Complexity} \citep*{Weaver1948}  which casts the nature of scientific problems into three classes: \emph{``problems of simplicity''}, described and modelled using simple sets of variables and equations that behave predictably; \emph{``problems of disorganised complexity''}, where the behaviour of large quantities of elements such as gas molecules in a container are modelled collectively through statistical methods even though the behaviour of each constituent part is chaotic; and, heralding the complexity sciences, \emph{``problems of organised complexity''}, which do not adequately yield to either of the approaches above and present some difficulty in solving because they exhibit non-linear, emergent, and adaptive behaviours. She places the nature of cities squarely in the last category --- problems of organised complexity --- entailing \emph{``a sizable number of factors which are interrelated into an organic whole''} and present \emph{``situations in which a half-dozen or even several dozen quantities are all varying simultaneously and in subtly interconnected ways''} \citep[][p.2]{Jacobs1961}. Attempts at planned settlements tend to eschew complexity in favour of, per Garden Cities, dumbed-down linear ratios between abstract quantities such as housing and jobs or, as symbolised by Le Corbusier's Radiant City, conceptions of urbanism rooted in larger-scale abstractions wherein people are reduced to statistical aggregations, again treated as simpler linear combinations of variables. By recasting cities through a reductionist lens, planners had come to make decisions that were detached from the complexities of functioning neighbourhoods and treated inhabitants as simplistic aggregations tantamount to \emph{``grains of sand, or electrons or billiard balls''} \citep[][p.437]{Jacobs1961}. Jacobs proceeds to offer a prescription: our understanding of cities should instead develop out of \emph{``the microscopic or detailed view\ldots rather than on the less detailed, naked-eye view suitable for viewing problems of simplicity or the remote telescopic view suitable for viewing problems of disorganised complexity''} \citep[][p.439]{Jacobs1961}. She outlines three principles to this end: first, think about city `processes': elements within cities have different effects depending on their combinations with other elements and the varied interactions between them; second, reason from an inductive rather than deductive approach, meaning from particularities to generalisations instead of from generalities to particulars; third, look for `unaverage' clues such as peculiarities, outliers, or nascent trends that help elucidate the workings of cities rather than fixating on statistical methods rooted in large-scale `averages' which may offer little explanation for how constituent elements may be operating within a complex system.

A knee-jerk reaction may be to reject machine learning out-of-hand for its links to mathematics and statistics more generally; however, on closer scrutiny, the synthesis of spatially precise urban morphological metrics combined with machine learning methods affords the use of highly detailed datasets capable of capturing and preserving contextual particularities; facilitates the use of high-dimensional datasets with significant assortments of variables and potentially complex and varied non-linear relationships between them; and, in the form of unsupervised methods combined with deep neural networks, allows for structures to be unearthed directly from within the data without the imposition of reductionist theories or formulas. In contrast to traditional statistical methods applied to larger-scale spatial aggregations, unsupervised machine learning methods applied to high-resolution and contextually-anchored spatial data resembles an approach more akin to proceeding from the particular to the general. Despite the large volumes of information, the data-space is (in effect) explored `line-by-line with model \emph{losses} computed and updated over comparatively small batches of data. Patterns are `sniffed out' using exploratory and bottom-up-like procedures with the more prevalent of these congealing over successive iterations to reveal thematic patterns that have arisen directly from the data. Emphatically, this reasoning only holds if working with pedestrian-scale metrics gathered using sufficiently high-resolution analysis, with the measures processed directly from each location. Use of intervening levels of spatial aggregation, interpolation from larger to smaller units of scale, or overly large units of analysis would otherwise result in the attrition of information and, critically, discards or otherwise masks local-scale inter-relationships between the variables.

Traditional forms of urban morphological analysis have been challenging to apply at scale because of reliance on manually collated observations and wearisome calculations. Geographic Information Systems (GIS) have permitted larger scales of quantitative analysis. However, the lack of comprehensive and granular data sources combined with computational constraints meant that these methods have oft been applied against larger units of spatial aggregation and relied on simplified distance metrics \citep{Logan2017, Araldi2016}. More recently, however, the increased availability of detailed datasets has facilitated a finer scale of analysis while retaining the ability to process larger areal extents \citep{Araldi2019}, thus prompting the adoption of multi-variable and multi-scalar workflows. The ensuing large and high-dimensional datasets can be combined with unsupervised exploratory methods and have engendered interest in how urban morphological analysis can be applied not only to the exposition of existing cities but also in the capacity of a rigorous design-aid for newly planned forms of development \citep{Serra2016, Gil2009, BerghauserPont2017, BerghauserPont2019}.
\section{Dimensionality reduction with PCA}\label{dim-red-pca}

The ever-increasing size, resolution, and dimensionality of urban datasets echoes explosive increases in data availability more generally. These developments have catalysed an assortment of related technologies and buzz-words: big data and the cloud, the internet of things and ubiquitous computing, and, not least, machine learning (ML) and artificial intelligence. The rapid adoption of data science and ML methods across a wide variety of fields is further fuelled by improved access to powerful forms of computation and openly available toolsets such as \code{Scikit-learn} \citep{Pedregosa2011} and \code{Tensorflow} \citep{Abadi2016}. Large and complex datasets can now be explored in ways that were difficult or impossible with the use of traditional scientific approaches; yet, these methods are more nuanced than the bantering-about of buzz-words may imply. In contrast to the human mind --- capable of forming highly developed abstractions aiding generalisable inferences based on relatively small amounts of data garnered from varied sources --- ML methods such as deep neural networks require tremendous volumes of data and substantial computing power to learn useful patterns \citep{Sinz2019, Marcus2018}. Another complication is that datasets used in ML are not only large, in terms of the number of sampled points, but are also frequently high-dimensional: the number of features described per data point can easily number in the hundreds or even thousands, and the myriad relationships between these features may be complex, on one-hand, or redundant, on the other. This complexity presents a quandary to analysis: whereas it can be trivial to visualise, explore, and intuitively understand the juxtaposition of a handful of features, the feasibility of doing so diminishes once the number of features increases and inter-relationships become more complex. High dimensionality is not only problematic for the visualisation and exploration of data but also the underlying analysis because, though the performance of ML algorithms at first tends to improve with the addition of features, it begins to deteriorate once the number of dimensions becomes sufficiently large \citep{Trunk1979}. This situation arises because the available data becomes increasingly sparse as the dimensionality increases and leads to \emph{overfitting} due to the invocation of the \emph{curse of dimensionality}: the amount of data necessary for ML algorithms to adequately generalise increases exponentially upon the addition of features, with inevitable implications for computational complexity \citep{Geron2017}.

Many ML methods can be ascribed to two general classes: \emph{supervised} and \emph{unsupervised}. Supervised methods train ML models against known quantities or labels, with the intent that the model should be capable of generalisation to predict unseen instances accurately. Such models may be relatively accurate, assuming that sufficient data is available and that adequate validation and testing procedures have been followed. However, their internal workings can be notoriously opaque, as is the case of deep neural nets --- the proverbial machine learning black-box --- and do not necessarily contribute to more comprehensive theoretical constructs of how or why the predictions work. Unsupervised methods, on the other hand, involve the exploration of unlabelled data for purposes such as reducing high-dimensional datasets to lower-dimensional representations (\emph{dimensionality reduction}) and the organisation of the data into representative groupings (\emph{clustering}). These approaches offer an opportunity for structures inherent to the dataset to bubble up, potentially aiding the discovery of themes that can more easily be tied to domain-specific interpretations. Regardless of intent, unsupervised methods may be necessitated by the lack of labelled data because it is often not feasible to label instances for large datasets manually.

The dataset developed through the remainder of this text consists of approximately one million samples for Greater London, each of which contains 162 dimensions for a variety of centrality, mixed-use, land-use accessibility, and population measures derived from the \emph{Ordnance Survey} \emph{Open Roads}, \emph{Ordnance Survey} \emph{Points of Interest}, and \emph{Office for National Statistics} census datasets, with measures computed for pedestrian distance thresholds ranging from $100m$ to $800m$ from the point of analysis. These points are spaced at $20m$ intervals on the road network, with the respective measures calculated with the \code{cityseer-api} package \citep{Simons2021b} for each location and weighted by spatial impedances (network distances) to increase their sensitivity to the sampled location. There are no clear-cut descriptors (labels) for the individual points of analysis at this level of resolution: is the point located on a vibrant high street or the dead-end of a suburban cul-de-sac? Do throngs of people frequent its location, or is it nestled by parkland somewhere in the vicinity of Greater London's greenbelt? Whereas forms of labels may be available for larger spatial units of analysis, they are not available at the granular level and their utility would substantially degrade if projected or interpolated from larger geographic units to smaller and more context specific locations. Nevertheless, the use of unlabelled data processed through unsupervised methods of analysis is not necessarily a drawback: the discovery of themes or patterns as inherent characteristics of the data can actually be preferable to processes that shoehorn data into existing classifications: nuances may be elucidated, or interpretations uncovered, that are more representative of underlying structures \citep{BerghauserPont2017}.

Dimensionality reduction methods find ways to express a higher-dimensional dataset in a lower-dimensional space. The data consequently becomes easier to visualise and explore, and the distillation of complex higher-dimensional datasets into underlying patterns or themes helps frame the observed phenomena in more succinct terms that may have greater relevance to intuitively held conceptions. The stalwart of these techniques, dating back to 1901, is \emph{Principal Component Analysis} (PCA): higher dimensional data is projected into a lower-dimensional orthonormal subspace that best expresses the variance of the original dataset \citep{Pearson1901}. Whereas higher dimensional data may contain many highly correlated features, PCA, in effect, `collapses' these overlapping regions of data into a set of linearly uncorrelated variables. These are described as the principal components and are typically arranged in descending order such that the first principal component represents the most significant amount of variance followed by subsequent components representing successively smaller variances. Depending on the data, it is often possible to explain a substantial percentage of the original variance with only a handful of principal components, though this is not necessarily the case for complex datasets spanning numerous underlying generative factors. In the classic case, the process of PCA encompasses mean centring and variance scaling of the data, followed by finding the covariance matrix and then solving for the eigenvectors and eigenvalues from which the principal components are derived. In practice, a variety of procedures exist, often based on \emph{Singular Value Decomposition} or any one of a variety of niche implementations for specialised use-cases, e.g. incremental (memory efficiency), kernel (non-linear), and sparse variants \citep{Jolliffe2016, Hastie2013, Geron2017, Raskutti2004, Zou2006}.

\begin{figure}[htbp]
 \centering
 \includegraphics[width=\textwidth, keepaspectratio]{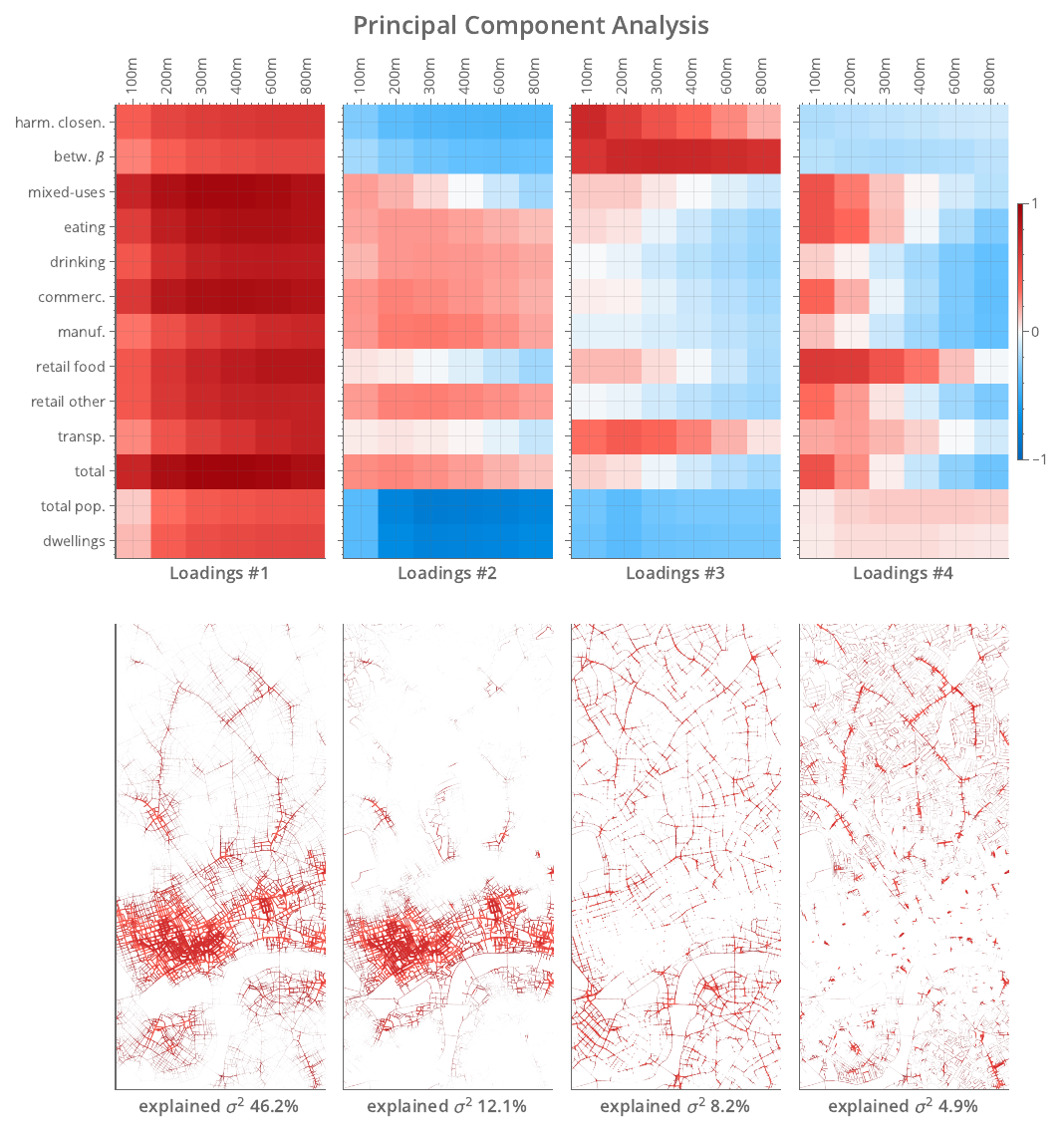}
 \caption[Loadings and spatial plots of PCA components.]{Loadings (correlations between the original variables and the latent component) and spatial plots of the first four Principal Components.}\label{fig:pca-b}
\end{figure}

Figure~\ref{fig:pca-b} shows the application of PCA (\code{Scikit-learn} SVD implementation) to the above-described dataset (see Table~\ref{table:dim-red-vars} for the input variables); the first Principal Component explains 46\% of the variance, drawing out commonalities across all themes and distances, with a particular emphasis on eating, commercial, and retail land-use accessibilities. Distances are primarily centred on mid to farther-distance pedestrian thresholds, indicating that these are district-wide urban areas instead of smaller-scale local neighbourhood clusters. This component accentuates areas such as the more lively parts of inner London and some intensive surrounding neighbourhoods such as Camden, Angel, and Dalston. These observations concur with generally held theories of complementarity between land-uses, centralities, and overall population densities and show that this is by far the most significant component. The second component explains 12\% of the variance and again focuses on inner London, but this time isolating mid-range concentrations of land-use accessibilities to the exclusion of populations and centralities. For emphasis, the PCA components are orthonormal, meaning that the variance described by this component is not a repetition of, or `overlap' with, the variances shown for the first component. The third component shifts away from central London, this time drawing out the street network and its relation to transportation accessibility and, to a lesser extent, local intensities of food-related retail. In turn, the fourth component accentuates high local concentrations of eating, commercial, and especially food retail locations as may be characteristic of neighbourhood high-streets and emblematic of the historical thoroughfares through London's many `villages'. The explained variance has now dropped to 5\%, with subsequent components becoming ever-more narrowly focused on a tremendous variety of themes. Supplementary Figure~\ref{fig:pca_components} shows the explained variance for the first twenty components, at which point a combined variance of 95\% is approached.
\section{Dimensionality reduction with Variational Autoencoders}\label{dim-red-vae}

A more recent counterpart to Principal Component Analysis is the \emph{autoencoder}, entailing the use of neural networks to learn a lower-dimensional \emph{internal representation} of a dataset \citep{Hinton2006}. There has been growing interest in the application of autoencoders (and PCA) to urban analysis, though this has tended to be in the context of visual forms of analysis where the autoencoder attempts to categorise and reconstruct rasterised representations of plans or photographs \citep{Moosavi2017, Law2019, Kempinska2019}. The following discussion and analysis are instead rooted in measures of network centralities, land-use accessibilities, and population densities assessed at a range of pedestrian walking thresholds, with the intention of unveiling patterns and relationships between these variables.

The underlying machinations of autoencoders, as with deep neural networks more generally, can be complex, but the intuition is relatively straightforward: the high-dimensional data is `squeezed' through a lower-dimensional informational bottleneck so that the autoencoder learns a succinct or `compressed' encoding for the original data. By learning to discern important features or patterns inside the reduced dimensional encoded space, the autoencoder can better recover the essence of the data once it is expanded back up to the original higher-dimensional form. Although autoencoders can be thought of as a form of \emph{unsupervised} learning, they are technically a \emph{self-supervised} process because the decoder's reconstructed outputs are evaluated against the original input data. The reduction of the original features to a lower-dimensional \emph{latent representation} (or `\emph{code}') is known as the \emph{encoding} step, whereas the expansion of the encoded data back up to the original number of features is known as the \emph{decoding} step.

Variational autoencoders (VAE) are the subject of interest in the proceeding discussion: VAE coerce the encodings to a continuous and normally distributed latent space (per dimension), thus conferring an advantage: VAE can be used as generative models because the latent space can be sampled at will to infer a congruous range of decoded outputs. Using a crude analogy, the latent dimensions behave like a set of adjustable dials capable of generating a range of decoded samples from underlying patterns in the data, as conveyed in Supplemental Figures~\ref{fig:sweep_1d} and~\ref{fig:sweep_2d}. VAE, and autoencoders more generally, are explained more comprehensively in Supplementary Section~2.

A VAE is here implemented in \code{tensorflow v2} \citep{Abadi2016} and is applied to the same data used in Section~\ref{dim-red-pca}. The number of latent dimensions is constrained to six to keep the visualisations in the subsequent figures to a more easily communicable level. Note that other forms of generative neural networks also exist, notably Generative Adversarial Networks \citep{Goodfellow}; however, these place emphasis on generative performance as opposed to the discovery of latent representations, and can be challenging to train and evaluate \citep{Hui}.

\begin{table}[htbp]
 \centering\footnotesize
 \begin{tabular}{ c c p{8cm} }
 Latent & average $D_{KL}$ & Description \\
 \midrule
 1 & 1.55 & Contrasts areas of higher centralities against areas of lower centralities. \\
 2 & 2.18 & The most dominant latent, emphasising collinearities between closeness centralities, mixed-uses, vibrant land-uses, and population densities. \\
 3 & 1.58 & Draws out areas of higher and lower population densities. \\
 4 & 1.16 & Distinguishes areas with high local intensities of food stores and transportation accessibility from areas with manufacturing. \\
 5 & 1.08 & Identifies areas with higher or lower access to transportation. \\
 6 & 1.11 & Separates areas with local centralities and population densities from locations with high concentrations of drinking establishments. \\
 \end{tabular}
 \caption{Summary of variational autoencoder latents for 6 dimensions.}\label{table:latents}
\end{table}

\begin{figure}[htbp]
 \centering
 \includegraphics[width=\textwidth, keepaspectratio]{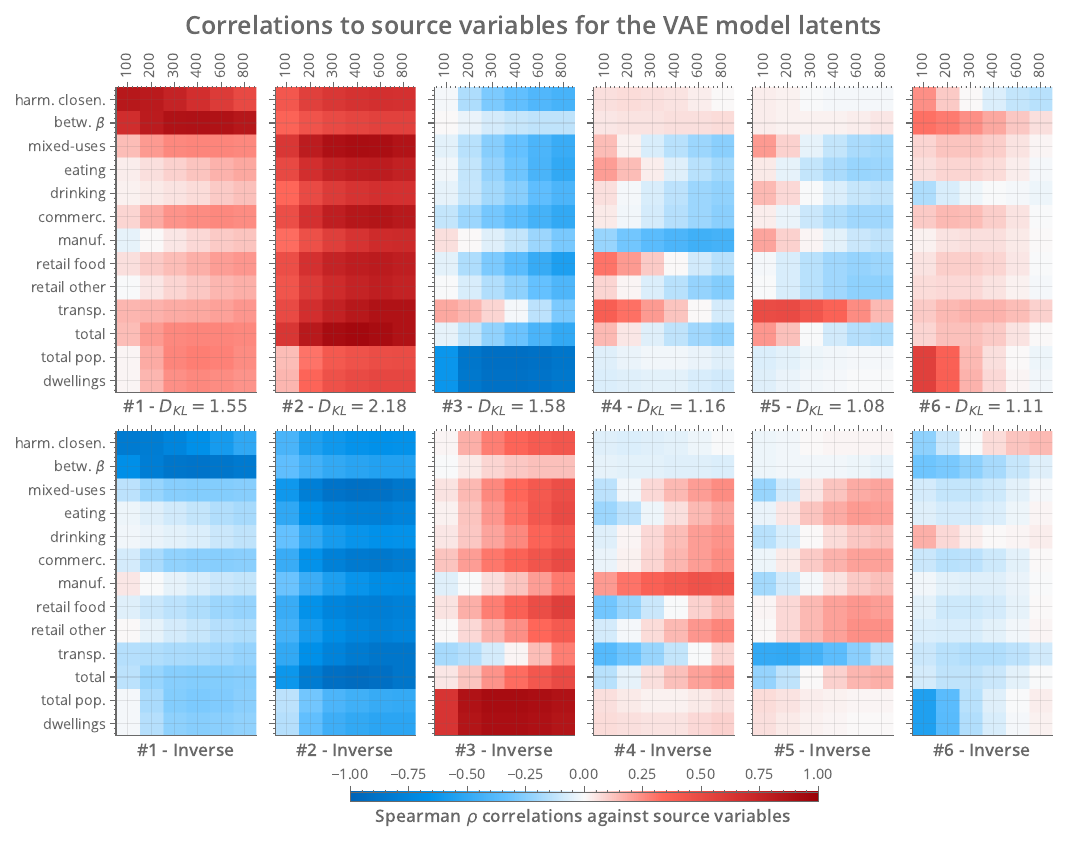}
 \caption[Correlations between VAE latent representations and source data.]{Correlations between latent representations and source data, with inverse correlations shown on the lower row.}\label{fig:latents_corr}
\end{figure}

Figure~\ref{fig:latents_corr} shows the correlations (against the original feature dimensions) and Supplementary Figure~\ref{fig:latents_map} shows geographic plots for each of the six latents. The more dominant latents assume the greatest amount of the available $D_{KL}$ capacity and, in the case of latents 1~\&~2, show some resemblance to the more dominant PCA components (compare Figure~\ref{fig:pca-b}). Table~\ref{table:latents} (compare Supplementary Figure \ref{fig:sweep_1d}) summarises the characteristics of each latent, revealing both similarities and differences when compared to PCA: in contrast to the essentially limitless number of principal components permitted by PCA, where it can be noted that the constraint on the number of VAE latent dimensions causes the VAE to develop more complex representations for the less dominant latents, such as focusing on particular combinations or contrasted pairings of land-uses, population densities, and centralities. 

\begin{figure}[htbp]
 \centering
 \includegraphics[width=\textwidth, keepaspectratio]{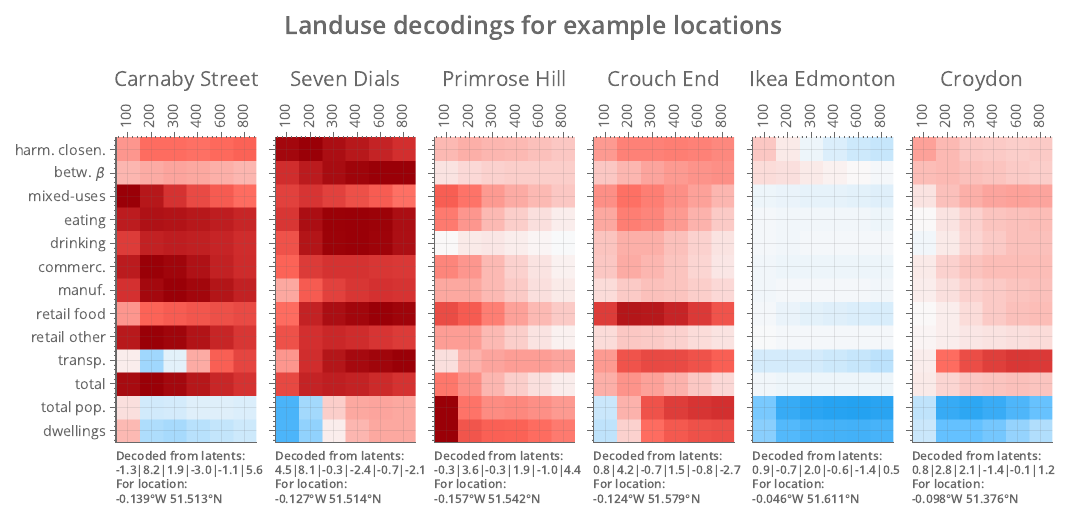}
 \caption[Example decoded latents for selected locations.]{Example encodings and correspondingly decoded latents for selected locations around Greater London.}\label{fig:example_latents}
\end{figure}

In practical terms, Figure~\ref{fig:example_latents} shows example latents and the corresponding decodings for a selection of locations in Greater London: Carnaby Street and Seven Dials are distinguished by exceedingly high values for latent 2, cementing Soho's status as an outlier in terms of local access to mixed-uses, eating, drinking, and retail land-uses, with Seven Dials further accompanied by exceptionally high closeness and betweenness centralities as reflected in latent 1. Primrose Hill is emblematic of a smaller-scale residential high-street as distinguished by latent 4, indicating high pedestrian access to local food retail locations, and latent 6, a high local population density. Whereas Primrose Hill shows a low degree of wider neighbourhood-scale intensities of land-uses and populations, Crouch-End is an example of a neighbourhood where these are present to a high degree. Ikea Edmonton is a study in contrasts, showing no relation to pedestrian-scale land-uses and a negative relationship to population density (the inverse of latent 3); likewise, Croydon Wellesley Road has low local population densities and poor local access to land-uses, but slightly better access to farther land-uses.

\section{Clustering with deep neural nets}\label{clustering}

Variational autoencoders aid in the exploration of underlying generative factors, and these can be disentangled and explored in isolation or across various combinations across a range of values. With clustering, on the other hand, the emphasis shifts to how characteristic themes present within the data can be used to group similar observations. Clustering has seen adoption in computational urban morphological analysis, where it is used to identify distinct groupings across multiple variables such as centralities, population densities, and building typologies. \citet{Gil2009} applies K-medoids in the comparison of two Lisbon neighbourhoods, allowing for their classification according to multiple selected attributes; K-medoids is likewise applied by \citet{BerghauserPont2017} in the analysis of betweenness centralities and population densities. K-means, which scales better to larger datasets, has been applied to the comparison of street, plot and building types across five cities \citep{BerghauserPont2019} and to the identification of street-network archetypes \citep{Serra2016}. \citet{Araldi2019}, on the other hand, argue for a two-step process, where the first step incorporates an explicitly spatial method in the form of \emph{Local Indicators of Network Constrained Clusters} (ILINCS) \citep{Yamada2010}, a network-based implementation of \emph{Local Indicators of Spatial Association} (LISA) \citep{Anselin1995}. This method identifies spatially contiguous hot and cold spots --- locations where local intensities diverge from the underlying distribution --- for each of the respective input variables. The second step applies a probabilistic Bayesian clustering method for grouping locations based on overlaps in the spatial patterns identified in the first step \citep[also see][]{Araldi2016}.

In keeping with the first of the strategies mentioned above, the proceeding analysis derives clusters from multi-variable input data (see Table~\ref{table:clustering-vars}) without recourse to an explicit spatial clustering step. Whereas this approach foregoes spatial continuity as an outright aim, the clustering assignments for individual data points still reflects larger spatial patterns. As with the second of the mentioned strategies, a probabilistic clustering approach is employed in the form of a Gaussian Mixture Model (GMM) instead of a hard-clustering approach such as K-means or K-medoids. K-means identifies clusters through an iterative optimisation process alternating between a step assigning each data point to the closest cluster centroid and another step subsequently updating centroid locations to reflect the new mean position of the accordingly assigned data points \citep{Hastie2009}. K-medoids is similar to K-means but bases cluster centres on actual data points instead of the mean of a cluster, thus making it more resistant to outliers. A drawback is that it requires the quadratic $N^{2}$ computation of distances between data points and therefore does not scale well to larger datasets. K-means and K-medoids can be well-suited to the assignment of linearly separable spherical clusters of similar sizes, but their utility can be hampered when applied to continuously varying or overlapping sets of variables. High-resolution and high-dimensional urban morphological datasets are challenging in this regard because they typically do not present obviously delineated groupings of data; for example, whereas locations of notably high or low network centralities may exist, so too does a continuous spectrum of other areas with intermediary levels of centrality. The same is true for mixed-uses and population densities, and the interaction of these types of variables across multiple dimensions and at various scales of measure means that the delineations are not clear-cut. The clusters are, further, not necessarily similar in size and could be stretched or `tilted' through different dimensions, leading to problematic assignments in cases where data points from nearby clusters get caught up in the strictly spherical boundaries of another. Although conceptually similar to K-means, GMM clustering offers more subtlety due to the probabilistic assignment of data points based on proximity to each cluster's gaussian distribution. GMMs alleviate the above-described issues --- they can handle `squashed' clusters or clusters of different sizes --- and introduce a shift in perspective away from the binary true/false of hard classifications towards a more nuanced probabilistic model describing the likelihood of a data point belonging to any particular cluster. A data point located near a cluster's centre point would therefore belong to that cluster with a probability approaching one and to every other cluster with a probability nearer zero. In contrast, another data point located midway between two clusters could be considered as equally likely to belong to either. When the multivariate gaussian distributions of each cluster are taken together, they provide density estimation from which potentially complex data distributions can be modelled and sampled \citep{VanderPlas2016}. Gaussian Mixture Models are trained using an \emph{expectation maximisation} algorithm in an iterative process resembling that used for K-means: for a model consisting of $K$ components, the expectation step assigns data point $x$ a \emph{responsibility} (weight) for the $k^{th}$ component $C_{k}$ according to the likelihood of the gaussian with weight $\phi_{k}$ mean $\vec{\mu}_{k}$ and covariance $\Sigma_{k}$. Subsequently, the maximisation step updates the weighted means and covariances of each cluster to reflect the current assignments for all data points \citep{BrilliantGMM, Hastie2009}.

The application of clustering methods to large and high dimensional datasets can be challenging on several fronts. Some clustering algorithms, such as full $N^{2}$ distance-matrix-based similarity methods (where $N$ is the number of samples), are computationally complex or memory intensive and do not scale well to large datasets. Other algorithms, such as feature-based $N\times D$ methods (where $D$ is the number of features), scale more suitably \citep{Jiang2016}. However, their performance may still deteriorate when applied to high dimensional datasets due to the \emph{curse of dimensionality}: the increasing sparsity of high dimensional data spaces causes distances between points to become more uniform, potentially confounding the distance metrics underpinning clustering algorithms. The application of clustering to reduced dimensional latent representations is, therefore, a tempting strategy, though it does entail the risk of collapsing subspaces within which specific clusters may exist \citep{Steinbach2004}. These considerations have stimulated interest in techniques affording the joint optimisation of lower-dimensional representations and clustering memberships. \citet{Xie2015, Min2018} propose unsupervised Deep Embedded Clustering (DEC) in which a neural network learns a mapping from the original feature space $X$ to a lower-dimensional space $Z$ in which the Kullback Liebler divergence $D_{KL}$ between soft clustering assignments and a self-training target distribution is iteratively optimised. \citet{Jiang2016}, in turn, propose Variational Deep Embedding (VaDE), another unsupervised method, which likewise jointly optimises a lower-dimensional representation in parallel with clustering assignments. In this case, the implementation generalises a variational autoencoder by replacing the canonically single gaussian prior with a \emph{mixture of gaussians}. This method offers the advantage of being both unsupervised and generative, and it is possible to generate samples from the learned representations, as is the case with standard variational autoencoders. Other variants of Gaussian Mixture Model interpretations of autoencoders have also been proposed \citep{Dilokthanakul2016, RuiShu2016}. Out of these models, VaDE was selected after exploratory implementations because it was not as difficult to train, less memory intensive, and gave good clustering behaviour capable of scaling to large multi-dimensional datasets. The number of clusters was constrained to 21 for purposes of visualisation; see Supplemental Section~3 for additional discussion on the clusters and for comparisons of VaDE against other forms of GMM.

\begin{figure}[htbp]
 \centering
 \includegraphics[width=\textwidth, height=0.975\textheight, keepaspectratio]{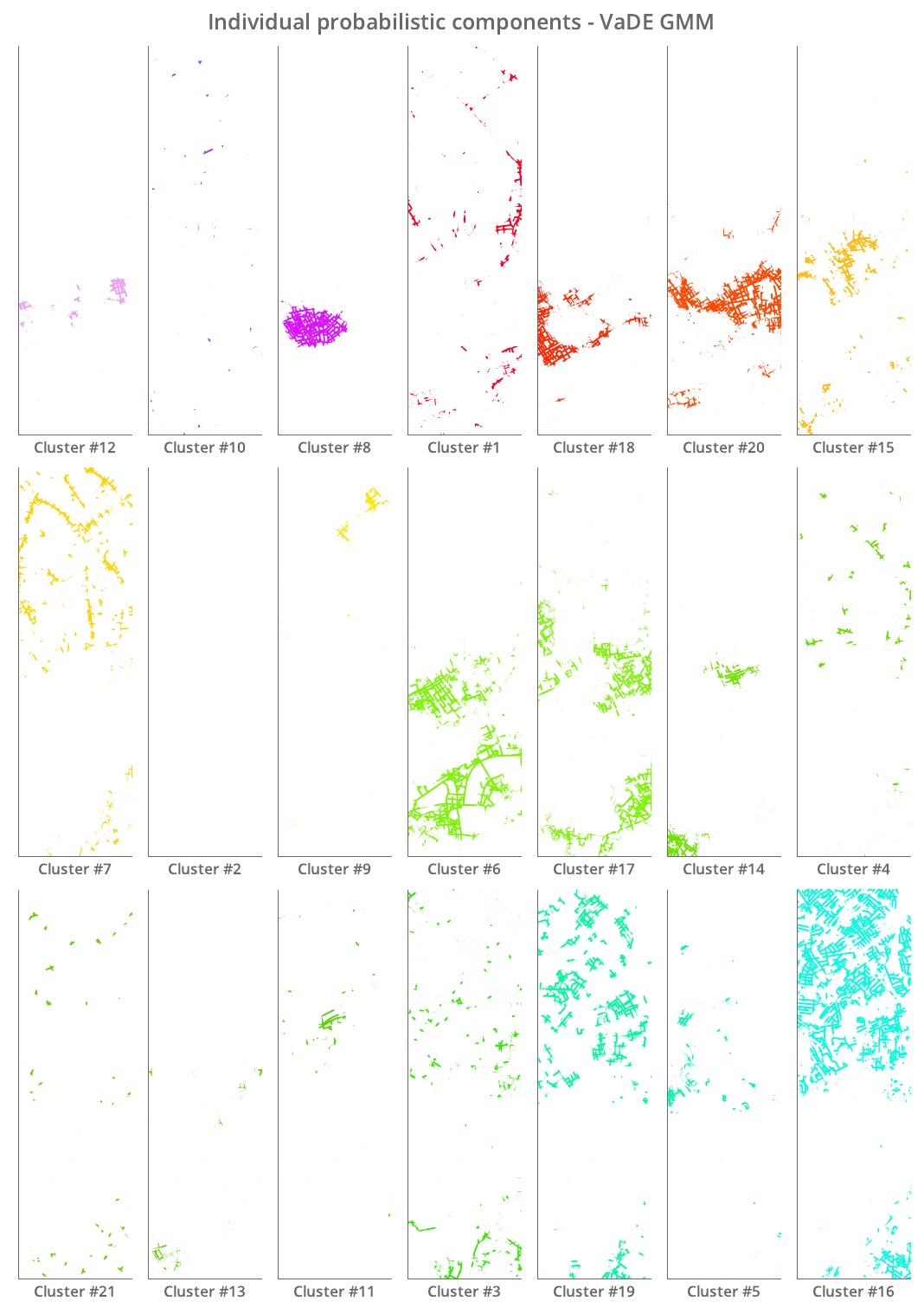}
 \caption[Individual VaDE cluster plots.]{Individual VaDE cluster plots, plotted in order of average local mixed-use vibrancy.}\label{fig:individual_clusters}
\end{figure}

\begin{figure}[htbp]
 \centering
 \includegraphics[width=\textwidth, height=0.975\textheight, keepaspectratio]{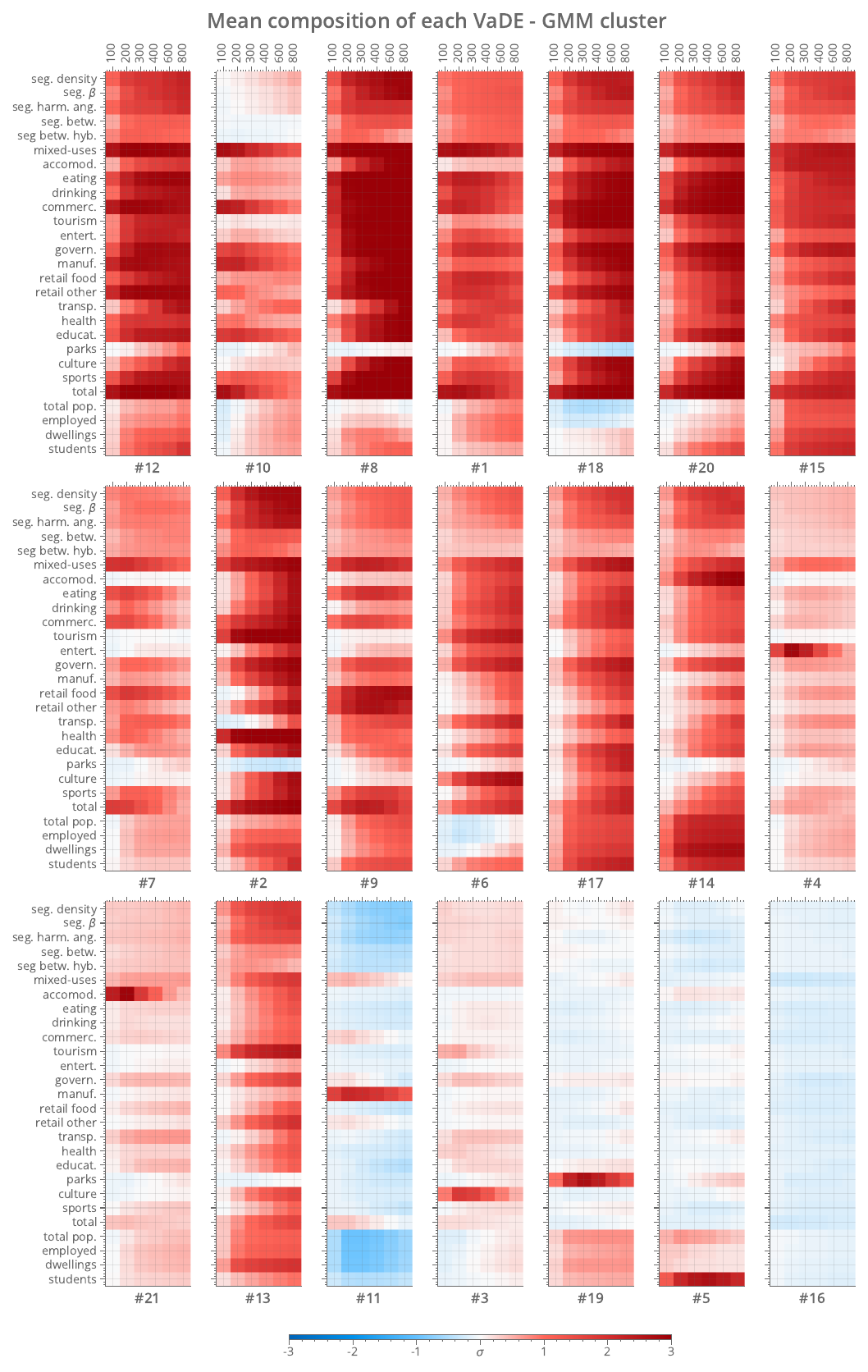}
 \caption[Mean values for each VaDE clusters.]{Mean values for each VaDE cluster.}\label{fig:cluster_mus}
\end{figure}

\begin{figure}[htbp]
 \centering
 \includegraphics[width=\textwidth, height=0.85\textheight, keepaspectratio]{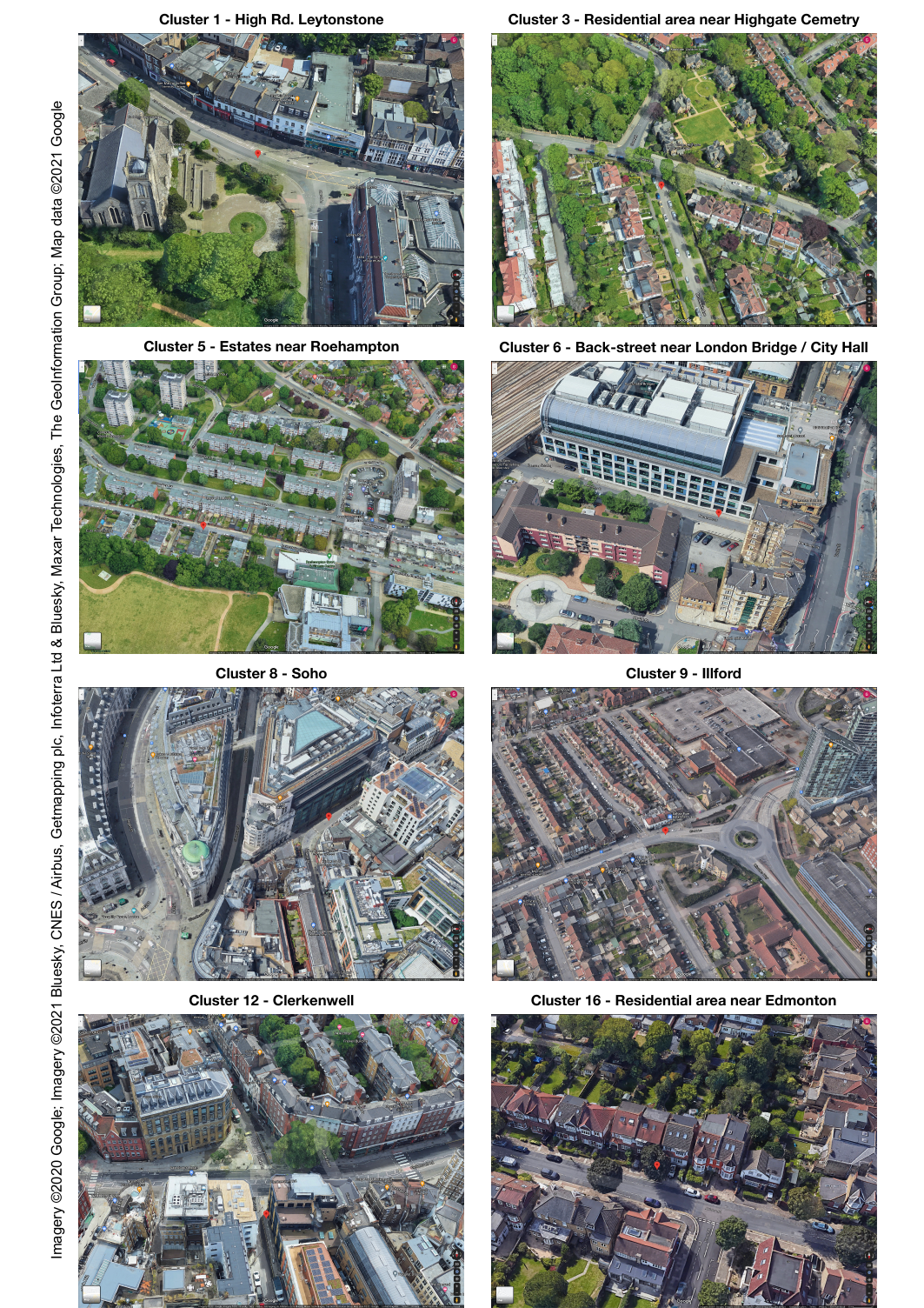}
 \caption[Example locations for selected VaDE clusters.]{\scriptsize{Example locations for eight selected clusters: selected based on the data point with the smallest norm to a cluster's mean. Cluster \#1: localised intensities of land-uses typical of high-streets, with a weaker emphasis on larger distance accessibilities; Cluster \#3: weak to middling access to most land-uses (typical of residential areas) but pronounced access to cultural (highgate cemetery and environs) combined with relatively low population densities; Cluster \#5: weak access to land-uses but high student population densities (Whitelands College and University of Roehampton); Cluster \#6: high access to medium-distance land-uses but less pronounced access to land-uses in the immediate surrounds; Cluster \#8: extremely high access to land-uses and closeness centralities; Cluster \#9: access to high-street specific land-uses (eating establishments on Ilford Lane to the left) but lower access to other land-uses; Cluster \#12: High access to land-uses and centralities; Cluster \#16: Nondescript centralities, land-uses, and densities.}}\label{fig:cluster_egs}
\end{figure}

Individual components are plotted in Figure~\ref{fig:individual_clusters} with a geographic plot of all clusters provided in Supplemental Figure~\ref{fig:VaDE_clusters}; average feature values of each cluster are plotted to Figure~\ref{fig:cluster_mus}; and a selection of cluster examples are plotted to Figure~\ref{fig:cluster_egs}. VaDE benefits from the joint optimisation of a lower-dimensional representation in concert with the assignment of clustering memberships. Visual inspection shows that the clusters have become more focussed on both inner London and neighbourhood high streets; for example, the most vibrant areas of inner London (\#12, light pink) and Soho (\#8, pink) are clearly delineated; other pockets of vibrancy, such as Camden, Angel, Kings Cross, Dalston, and Brick Lane are identified (\#1, deep red); high streets and ancillary areas are now encapsulated by degrees of intensity (\#1, deep red; \#7, yellow; \#4, deep green); and several thematic groupings emerge, such as manufacturing, parks, student population, entertainment, etc. Clusters focusing on inner London (e.g. \#8, \#12, \#18, \#20) tend to accentuate larger walking thresholds ($400m$ to $800m$) whereas those focussed on high-streets and pockets of intensive land-uses away from inner London focus on smaller thresholds ($100m$ to $400m$).

\section{Summary}

It is encouraging that the application of unsupervised dimensionality reduction to large datasets broadly encapsulating concepts of centrality, mixed-uses, land-use accessibilities, and population densities at pedestrian walking thresholds yields observations concurring with generally-held conceptions of urban theory. Significantly, the patterns that emerge go beyond simple mantras to expose tremendously rich and nuanced structures within the data that make sense upon reflection. For example, different kinds of land-uses cluster together in different types of scenarios and can move either in isolation or in lockstep with other classes of measures.

Assuming the use of a pre-packaged implementation such as \code{Scikit-learn}, Principal Component Analysis offers a quick and comparatively simple method to reduce the dimensionality of the data and uncover prominent themes. It can be an indispensable exploratory tool for large multi-feature and multi-scalar datasets, regardless of whether followed by more esoteric methods. Autoencoders entail a substantial increase in complexity in terms of understanding their workings and implementary specifics. These cannot typically be used in a simple `off-the-shelf' manner and entail a non-trivial degree of proficiency in the use of programming languages and an understanding of machine learning packages such as \code{Tensorflow} so that ML models can be built and correctly trained. Beyond the additional investment in time and complexity, autoencoders offer a tremendously powerful toolset for exploring large datasets in a manner that can be moulded to specific lines of inquiry. Autoencoders --- particularly of the variational kind --- provide decipherable glimpses of their inner workings and can be employed as powerful generative models rooted in structures and themes learned from within the data.

Clustering methods such as K-means and Gaussian Mixture Models scale relatively well to large datasets and allow for data points to be grouped by their characteristics. These groupings provide an opportunity to elucidate themes uncovered within the data and help bridge the quantitative realm of raw metrics and their mathematically abstract machinations to more relateable qualitative conceptions that can be relayed to urbanists more generally. In essence, observations starting as metrics for a particular point in space --- a jumble of numbers representing network centralities, mixed-uses, land-use accessibilities, and population densities --- can more easily be articulated in the form of a concept such as a `vibrant inner-urban area', `high street', or `quiet residential area' than to explain the meaning of Hill Numbers or the encoded latents of an autoencoder.

Whereas powerful, these methods also deserve some caution: the validity of unsupervised methods is dependent on the quality of the data: spurious observations may arise if applied to non-standardised variables; coarse or inferior quality data sources; or if excessively broad interpretations are ascribed to limited datasets. There can also be a tendency to interpret observations in absolutist terms. This perception should be guarded against when conveying these methods and the ensuing conclusions to urban designers, architects, and planners who might not come from a computational or data-science background. Put differently, conclusions derived from unsupervised methods are inherently connected to the underlying data and can therefore be more fluid than the impressions sometimes given. For example, changing how metrics are measured, using different groupings of measures, using different distance thresholds at which measures are calculated, or changing the form of analysis --- such as how data is preprocessed, the number of latent dimensions, or the number of clusters --- will all cause the themes within the data to morph accordingly.

There remains a need for further research to elucidate how visual analytic tools can be developed to more fully exploit the explanatory power of these methods in a manner that is not overwhelming to human cognition.

%% supplementary section
\setcounter{figure}{0}
\makeatletter 
\renewcommand{\thefigure}{S\@arabic\c@figure}
\makeatother

\setcounter{table}{0}
\makeatletter 
\renewcommand{\thetable}{S\@arabic\c@table}
\makeatother

\clearpage
\section{Supplementary Material}
\subsection{Supplementary Section 1: PCA}

\begin{figure}[h]
  \centering
  \includegraphics[width=\textwidth, keepaspectratio]{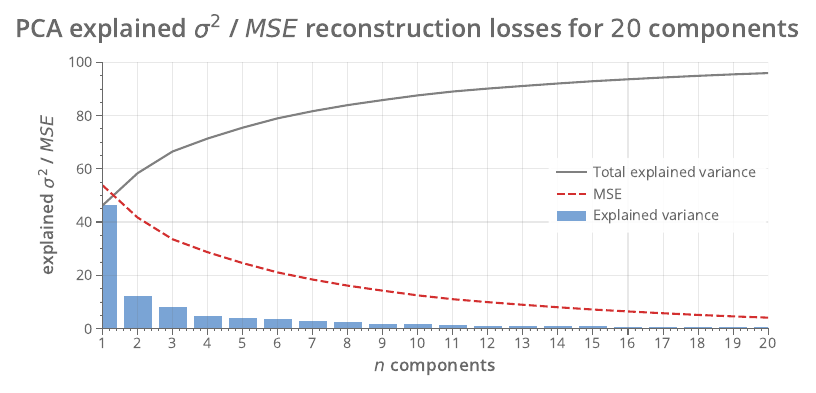}
  \caption[Figure S1: Explained variance and $MSE$ for the first twenty PCA components]{Explained variance / $MSE$ reconstruction loss for the first twenty PCA components.}\label{fig:pca_components}
 \end{figure}

\subsection{Supplementary Section 2: Autoencoders and VAE}

Neural networks typically consist of an input layer, an output layer, and one or more hidden layers through which the data passes. Each layer consists of several \emph{neurons}, each of which employs an \emph{activation function} to determine how and when this neuron `fires' to pass information to the next layer. In the case of a standard \emph{fully connected} network, each neuron is connected to every neuron in the previous layer, and these connections are weighted. Consequently, the signal ultimately reaching any particular neuron is a combination of all of the weighted inputs gathered from the neurons in the preceding layer (which, in turn, are all individually connected to every node in the proceeding layer, etc). Each neuron also has a bias parameter, thereby controlling the threshold at which the activation function will trigger. Once the forward pass of the data has proceeded through all layers from input to output, a \emph{loss function} (such as \emph{mean squared error} or \emph{cross entropy}) determines the accuracy of the output and the loss can then be differentiated backwards through the network using the \emph{chain rule}. This process is known as \emph{back-propogation} \citep{Rumelhart1987} and determines the extent to which each neuron contributed to the loss, after which a gradient descent optimisation algorithm updates the weights for each neuron. Weights for deep neural nets are ordinarily randomly initialised using methods such as \emph{glorot uniform} initialisation \citep{Kumar2017}: forward and backwards passes are then repeated until the weights converge on a solution minimising the loss function.

Autoencoders are, in effect, two neural networks --- the encoder and decoder --- sandwiching the reduced dimensional latent representation with the loss calculations and gradient descent applied to the model as a whole. Interestingly, if the network were to be optimised using purely linear activation functions, it would learn a latent representation resembling Principal Component Analysis \citep{Geron2017}. In practice, linear activation functions defeat the purpose of neural nets because these would otherwise collapse into a simplified linear combination of the input variables. For this reason, activation functions ordinarily assume a differentiable non-linear form such as \emph{ReLU} and its variants \citep{He2015}, and it is this capacity for non-linearity that allows neural networks to learn potentially complex patterns in the data. Another subtle but essential distinction between PCA and autoencoders is that PCA decomposes the data into a potentially large number of principal components, most of which are then ignored (by the user) in favour of selecting components explaining the bulk of the variance. On the other hand, autoencoders are configured with a preset number of latent dimensions, and the autoencoder will then attempt to use the available dimensions to full effect. The more latent dimensions, the better the quality of the reconstructions but the harder it becomes to visualise these dimensions and the more likely it becomes that the model may overfit.

Variational Autoencoders (VAE) are distinguished by a probabilistic approach: originally proposed by \citet{Kingma2013}, a given data point $x$ is encoded to a distribution over the range of possible latent parameters $z$ from which the data point may have been generated. Conversely, any given latent encoding $z$ maps to a distribution of possible decoded values of $x$. In contrast to conventional autoencoders, which directly map input variables to an encoding as a point in the latent space, VAE map each point to a mean $\mu$ and standard deviation $\sigma$ defining a gaussian distribution from which the encoding is randomly sampled. The use of a stochastic sampling step presents a barrier to back-propagation, so the \emph{reparameterisation trick} is applied to remove the noise variable $\epsilon$ from the chain of differentiation, resulting in the form $z = \mu + \sigma \odot \epsilon$ where $\epsilon$ is randomly sampled from a gaussian distribution $\epsilon \sim \mathcal{N}(0,1)$. The model loss is computed as the sum of the reconstruction loss (cross-entropy or $MSE$, as per conventional autoencoders) between the input and output, and the Kullback-Leibler $D_{KL}$ divergence of the latents between a gaussian prior $p(z)=\mathcal{N}(0,I)$ and the gaussian posterior approximation $q_{\phi}(z|x)$. $D_{KL}$ divergence is computed per:
\begin{equation}\label{eq:kl_loss}
 D_{KL} (q_{\phi} (z)||p_{\theta} (z)) = \frac{1}{2} \sum^{J}_{j=1} \big(1 + \log ((\sigma^{(i)}_{j} )^{2}) - (\mu^{(i)}_{j} )^{2} - (\sigma^{(i)}_{j} )^{2}\big)\, ,
\end{equation}
where $J$ represents the latent dimensionality of $z$; $j$ represents the $j$-th dimension of $J$; $\mu$ and $\sigma$ represent the variational mean $\mu$ and standard deviation $\sigma$; $i$ represents the $i$-th data point; and $\phi$ and $\theta$ represent the model parameters (weights and biases) for the encoding and decoding steps. See Appendix B of \citet{Kingma2013} for the full derivation.

A side-effect of the inclusion of $D_{KL}$ in the loss term is that the fidelity of the reconstruction will deteriorate as a consequence, which is a worthwhile and somewhat intentional tradeoff given that the motivation for using VAE is to uncover structures within the data. To this end, \citet{Higgins2017} introduce the $\beta$-VAE model, which adds a $\beta$ parameter to further amplify the contribution of the $D_{KL}$ term to encourage the \emph{disentanglement} of the latent dimensions ($\beta$-VAE). Disentangled representations attempt to isolate generative factors to a single dimension; for example, a representation of an object moving on a plane would be disentangled if movement on the $x$ axis is confined to one latent dimension and movement on the $y$ axis is confined to another. In contrast, entangled representations blur the contributions of independent factors, which may consequently be entwined with multiple other factors and spread across multiple latent dimensions. By more strongly penalising the Kullback-Leibler $D_{KL}$ divergence, the $\beta$ parameter imposes a more stringent limit on the capacity of the latent information channel: where $\beta=1$ corresponds to the original VAE model, $\beta > 1$ are associated with stronger disentangling because the model is encouraged to learn a more efficient representation of the data within the reduced capacity of the informational bottleneck. However, it should be noted that excessively high values of $\beta$ will cause a deterioration in the quality of disentanglement. The tradeoff, once again, is that the more strongly penalised $D_{KL}$ divergence causes yet further deterioration in the quality of the decoded reconstructions because the stronger information capacity constraint leads to a loss of high-frequency details. \citet{Burgess2018} consequently propose the addition of a capacity term $C$ to relax the information bottleneck constraint during training, leading to improved reconstructions while retaining the benefits of better disentanglement. The model loss consequently takes the form of:
\begin{equation}\label{eq:kl_cap}
 \mathcal{L} (\theta,\phi;x,z,C) = MSE + \beta\ | D_{KL} (q_{\phi} (z)||p_{\theta} (z)) - C |\, .
\end{equation}
$C$ is set to zero at the commencement of training and is ramped with each passing epoch until attaining its full value upon reaching the final epoch. Therefore, the model is encouraged to commence in a disentangled state, and the subsequent relaxation of the information constraint permits the accrual of additional information along the established trajectory. Whereas VAE are known to disentangle, there is no explicit theoretical basis explaining why the learned representations align well with the coordinate axes, and this has caused some bafflement amongst researchers. \citet{Rolinek2018} have subsequently shown that this behaviour can be attributed to `accidental' side-effects of canonical implementations: VAE exploit differences in variance to form latent representations in a manner resembling PCA (in the linear case), an indirect consequence of the enforced diagonal posterior of the encoder combined with the optimisation of the stochastic loss term, resulting in the orthogonality of the internal representations.

The degree of disentanglement is sensitive to the choice of hyperparameters $\beta$ and $C$. Whereas it is possible to gauge the general effectiveness of disentanglement through visual comparisons of parameter sweeps across selected latent representations, this approach lacks scalability and is subject to a degree of interpretation. Supervised disentanglement metrics are available for use in cases where sample classifications already exist, but these metrics can be inconsistent even if applied to the same data and for the same hyperparameters. This work proceeds with the use of Unsupervised Disentangled Representation (UDR). Giving the example of different colour systems, \citet{Duan2019} argue that the reliance of unsupervised metrics on ground-truth labels of canonical generative factors (in cases where these are available) is in and of itself subject to interpretation: for example, should colour be described in terms of HSV, HSL, RGB, or another colour space? They propose an unsupervised metric, Unsupervised Disentangled Representation (UDR), which takes advantage of an important aspect of disentangling: disentangled models tend to converge to similar disentangled representations in contrast to entangled models (and neural nets in general), which can converge to starkly different internal representations even if otherwise comparable in terms of data, hyperparameters, and performance. For each hyperparameter setting, UDR compares sets of models initialised with different random seeds: a pairwise matrix of correlations is computed between the latent dimensions of each pair of models; correlations are cast to positive values; the square of the strongest correlation is divided through the sum of all correlations, and the mean across the matrix is then returned. UDR indicates how well a set of models aligns while considering that the same generative factors may have been encoded to different latent dimensions or an inverse sign. 

\begin{figure}[h]
  \centering
  \includegraphics[width=\textwidth, keepaspectratio]{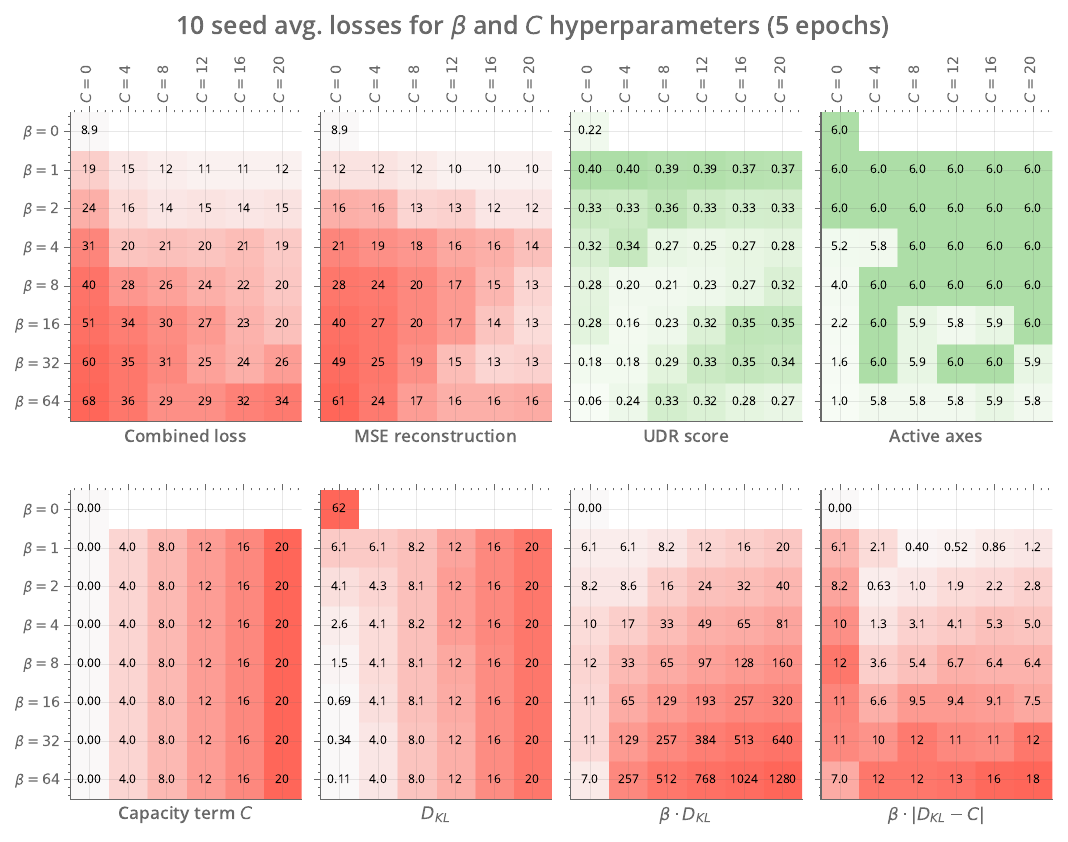}
  \caption[Figure S2: VAE losses for a range of $\beta$ and $C$ parameters.]{VAE validation losses (20\% split) and metrics for a range of $\beta$ and $C$. Best score of 5 epochs averaged over 10 seeds of random weights initialisation.}\label{fig:grid_search_fine}
 \end{figure}

Figure~\ref{fig:grid_search_fine} shows the effect of various strengths of $\beta$ and $C$ parameters for the shown validation losses and metrics. $\beta=0$ nullifies the contribution of the $D_{KL}$ term in the loss function, thus reducing the VAE to a conventional autoencoder. Whereas this results in the lowest reconstruction loss, it also yields the highest $D_{KL}$ due to the lack of regularisation of the latent space. $\beta=1$, in turn, corresponds to the base form of variational autoencoder (where $\beta$ is exactly 1), showing a corresponding increase in the reconstruction loss against a substantial decrease in $D_{KL}$, with the trend subsequently continuing for stronger levels of $\beta$. The addition of the $C$ term provides the anticipated improvement to the reconstruction losses while allowing the $D_{KL}$ to climb within the permitted extents. Once $\beta$ reaches 4, the heavier penalisation of the $D_{KL}$ loss becomes sufficiently significant to turn off some of the available latent axes due to the limit of the informational capacity of the latent space. As the capacity term $C$ is relaxed through training; the additional informational capacity increasingly allows the activation of all axes.

Variational autoencoders show a marked improvement in disentanglement for the given dataset compared to autoencoders without a variational component ($\beta=0$). For this analysis, where the number of latents is constrained to six, $\beta=1$ and $C=8$ provide a tradeoff between the quality of disentanglement and the reconstruction loss. These values selected for the subsequent plots, but it must be emphasised that the most suitable values of $\beta$ and $C$ are affected by the data source and the number of available latent dimensions. VAE with additional latent dimensions permit a cleaner separation of generative factors and may benefit from larger values for $\beta$ and $C$. For example, eight latent dimensions improved the UDR scores, with the best results attained for $\beta=4$ and $C=12$. Cases of higher UDR values for yet higher values of $\beta$ are not necessarily desirable because this can happen for situations where some latent dimensions have become inactive due to the informational constraints imposed by the stronger $\beta$. A tangential observation is that only the more dominant latents tend to remain activated in these situations and that these tend to be consistent across various values of $\beta$ and $C$.

 \begin{figure}[h]
  \centering
  \includegraphics[width=\textwidth, keepaspectratio]{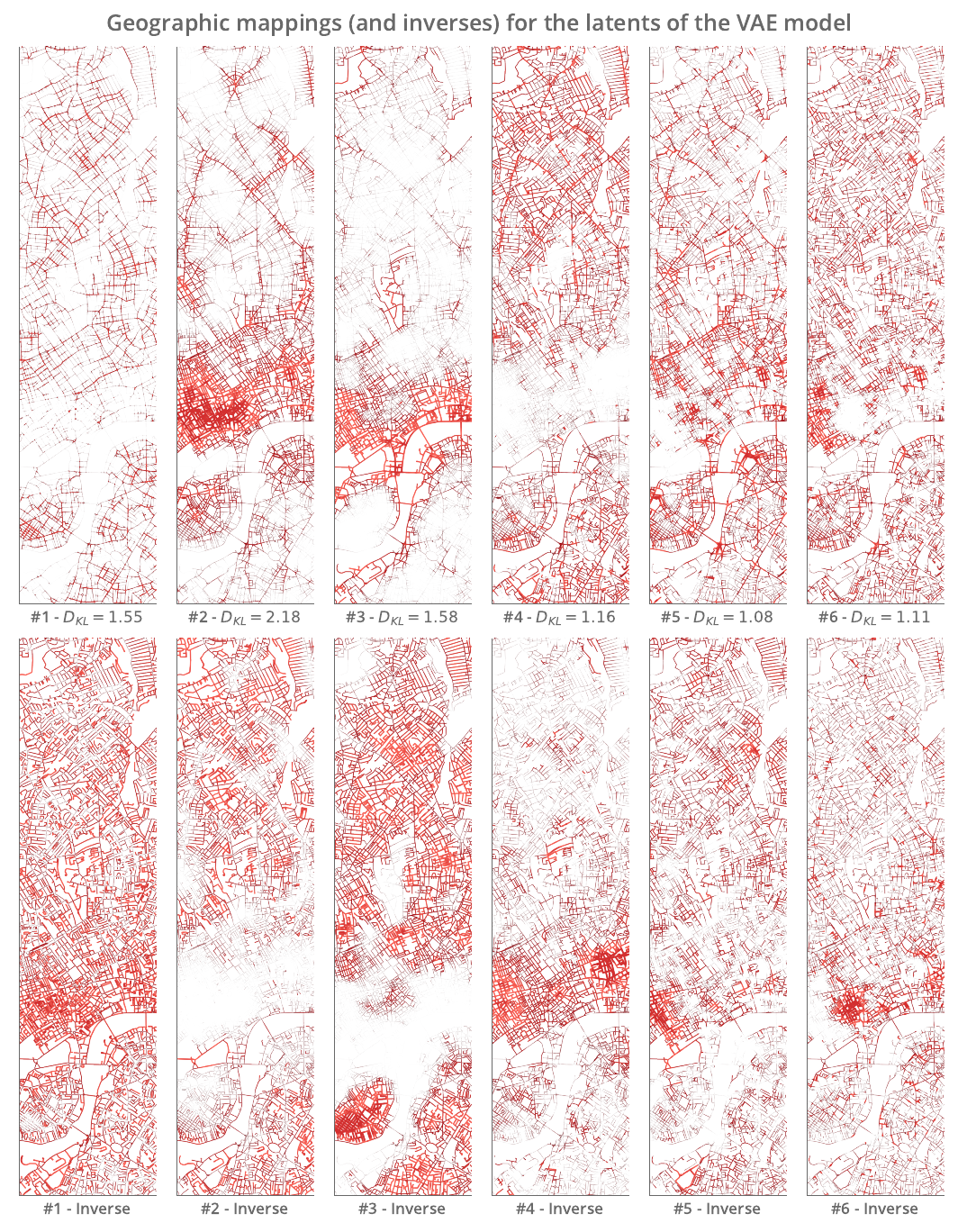}
  \caption[Figure S3: Spatial plots of VAE latent representations.]{Spatial plots of VAE latent representations, with inverse plots shown on the lower row.}\label{fig:latents_map}
 \end{figure}

\begin{figure}[h]
  \centering
  \includegraphics[width=\textwidth, height=0.975\textheight, keepaspectratio]{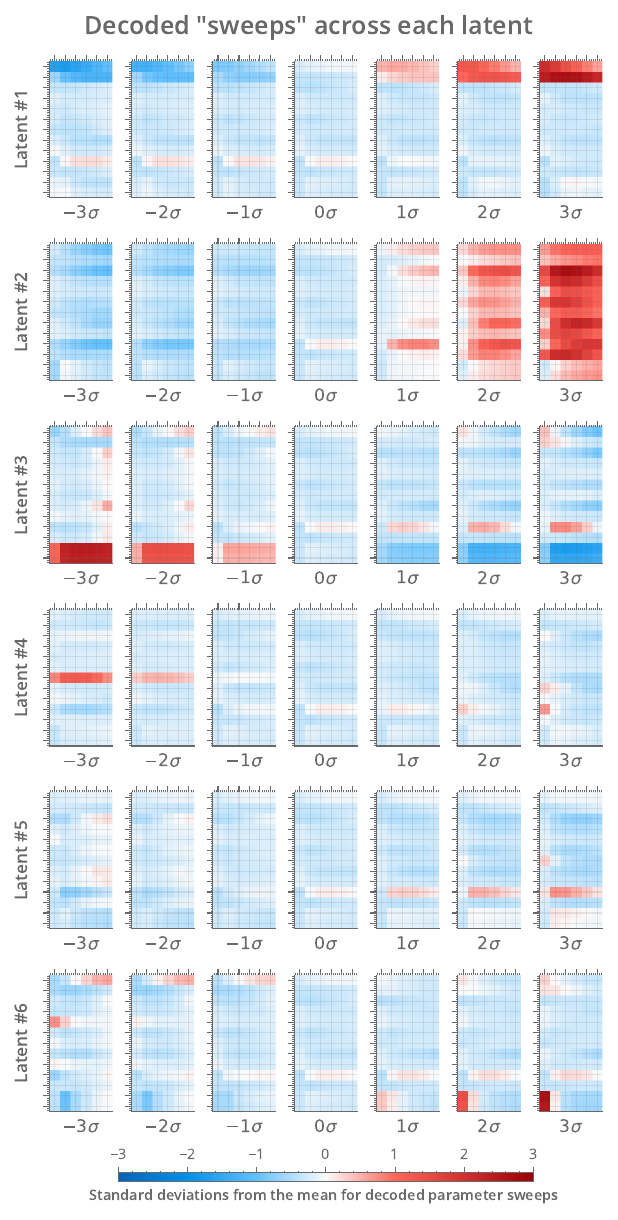}
  \caption[Figure S4: Sweep plots for the latents of the VAE model.]{Sweep plots showing decoded representations for each of the latents of the VAE model from $-3\sigma$ to $+3\sigma$.}\label{fig:sweep_1d}
 \end{figure}

\begin{figure}[h]
   \centering
  \includegraphics[width=\textwidth, height=0.975\textheight, keepaspectratio]{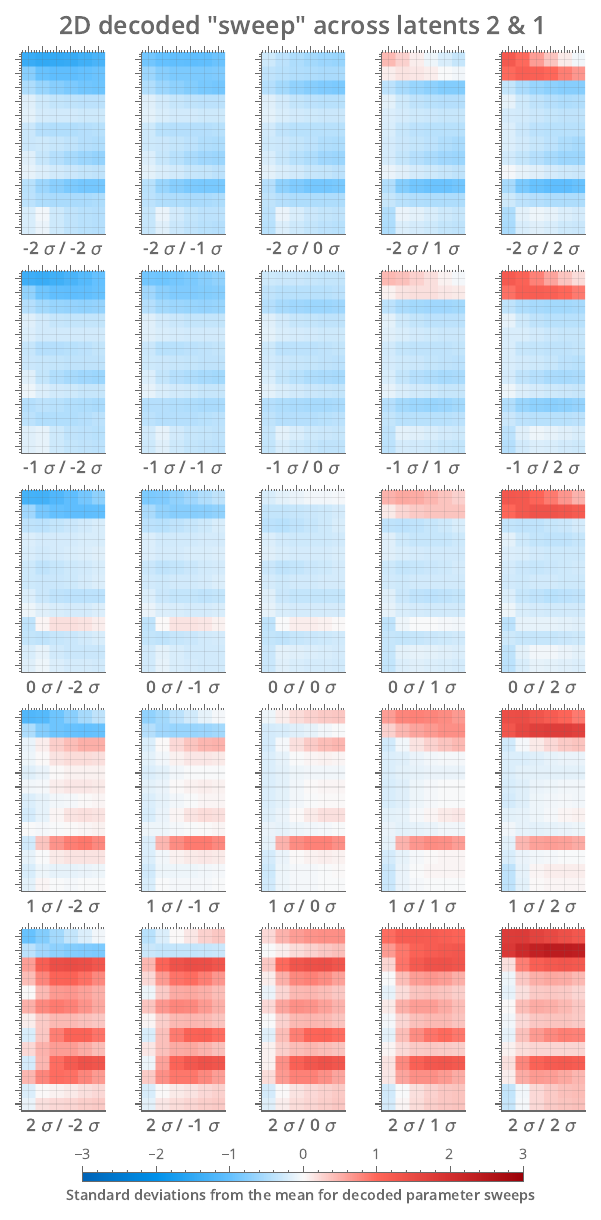}
  \caption[Figure S5: 2D sweep plot for (two) selected latents of the VAE model.]{2D sweep plot of decoded representations for latents \#1 \& \#2.}\label{fig:sweep_2d}
 \end{figure}

The continuous range of the encoded latent space becomes clearer in Figure~\ref{fig:sweep_1d}, which shows parameter sweeps for each latent dimension: a range of standard deviations, e.g.~$\{-3, -2, -1, 0, 1, 2, 3\}$, is decoded while holding all other latents steady at the mean. This yields a set of values spanning the encoded space of each latent; for example, the plot for latent 1 takes a range of encoded values varying from one extreme of~$z=\{-3, 0, 0, 0, 0, 0, 0, 0\}$ through to the other extreme at~$z=\{+3, 0, 0, 0, 0, 0, 0, 0\}$. The encoded values are then decoded and plotted to the figures, representing how a particular latent relates to the raw data in the original data space. The sweep plots provide a visual representation of how disentangling works: the variables described by a particular latent are free to vary from one extent to another while leaving unrelated variables as unaffected as possible (within the constraints of the number of latents and the $C$ parameter). Note that the plots spanning from negative to positive decoded latents are not necessarily symmetrical; autoencoders are capable of non-linear behaviour, which is observable to varying extents in some latent sweeps. Removal of the variational constraint (i.e. using a standard autoencoder) or an excessively large $C$ causes a deterioration of the disentangling and results in observably `messier' sweep plots. Whereas the respective latents gravitate towards disentanglement, it must be emphasised that they do not operate in isolation and that the encoded space can vary simultaneously across all available latents; per Figure~\ref{fig:sweep_2d}, the essence of this concept can be conveyed in two dimensions.

Whereas the latents of the VAE model encapsulate themes consisting of various combinations of centrality, land-use, and population density data, it is also interesting to consider how sub-themes within each of these categories might combine to inform the latents of the overarching model. To this end, this implementation consists of a modified encoding step that separates the neural network into three underlying encoding streams --- centralities, land-uses, and populations --- each of which is a self-contained sub-encoding step nested within the encompassing encoder. Consequently, the three channels learn category-specific internal representations of the data, which are then blended by the combined stochastic sampling step outputting the final latent representation. Therefore, the individual sub-latents allow these thematic structures to be uncovered while remaining under pressure from the combined model's loss function. The purpose of pipelining is to learn sub-latent structures with specific relevance to the combined context of the greater model. Whereas the respective themes (e.g. centralities or land-uses) could have been explored in complete isolation, the resultant structures might not necessarily be the same as those playing a role when considered in combination with other thematic streams. In the interest of brevity, these are not exhaustively unpacked in this discussion.

Experiments with convolutional layers and LSTM (long-short-term-memory recurrent) layers instead of `conventional' fully connected layers showed promise. They were, nevertheless, more fickle, slower to train, and sometimes yielded unintuitive latent representations, though they may be helpful in situations where clear benefit over conventional neural networks can be demonstrated. These were considered because the repetition of metrics across multiple distance thresholds allows for higher-level abstractions applicable to distance ranges. In the convolutional case, the data was reshaped by the number of features (y-axis) and the number of distance thresholds (x-axis), with the convolutional strides then applied per feature across multiple distances. The LSTM layers likewise applied the recurrences across the distance dimensions for the respective features.

\subsection{Supplementary Section 3: Clustering}

\begin{figure}[h]
  \centering
  \includegraphics[width=\textwidth, keepaspectratio]{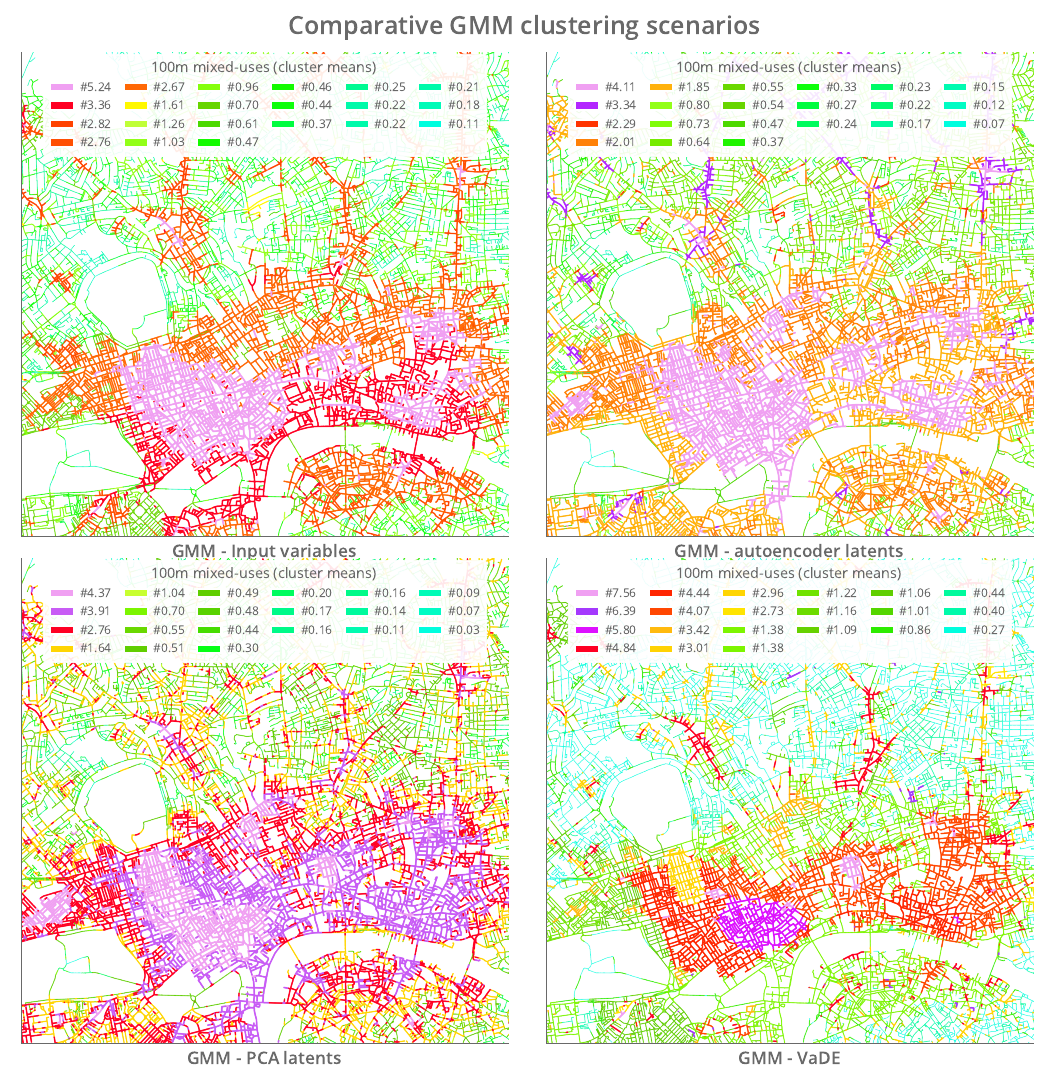}
  \caption[Figure S6: GMM comparisons.]{GMM clustering applied to four different scenarios: the full-dimensional feature space containing all data points and all 162 features; a reduced 32 dimensional latent space of an autoencoder; the principal components of PCA; a VaDE deep neural net implementation. The components are coloured according to the mean local mixed-use intensity.}\label{fig:gmm_comparisons_maps}
 \end{figure}

Supplemental Figure~\ref{fig:gmm_comparisons_maps} shows a comparison of GMM clustering (21 components) applied to four different scenarios: the first three implementations make use of a \code{Scikit-learn} GMM implementation \citep{Pedregosa2011}, whereas the fourth implements the VaDE deep neural net implemented in \code{tensorflow v2} \citep{Abadi2016}. Each cluster is coloured according to the group's mean local mixed-uses to aid visual identification and comparisons between the scenarios. In the first case, all 162 feature dimensions are fed directly to the GMM. The most vibrant cluster (light pink) identifies lively locations within inner London, such as Soho and Shoreditch. Subsequent clusters (pink, orange, yellow) correspond to groupings of successively less vibrant locations, thus capturing the gist of livelier areas and high streets in contrast to quieter residential and industrial areas. The second scenario applies GMM to the 8 dimensional latent space of an autoencoder. The third applies GMM to the principal components generated by PCA. These tend to offer similar results, though to varying degrees of delineation and with different emphases. The fourth scenario applies VaDE (32 latents) to all 162 feature dimensions.

Various clustering metrics exist for the purpose of selecting an optimal number of clusters, including measures such as the Silhouette Coefficient \citep{Rousseeuw1987}, computed from the mean intra-cluster distance and mean nearest-cluster distances, and the Calinski-Harabasz score \citep{Calinski1974} (also known as the Variance Ratio Criterion), computed from the ratio between within-cluster dispersion and the between-cluster dispersion \citep{Pedregosa2011}. These methods can prove helpful in datasets consisting of sufficiently clear delineations between samples where increases in the number of clusters can lead to a readily apparent threshold of diminishing returns, visible as an `elbow' or plateau when plotting the scores across a range of $n$ clusters. However, tentative exploration of these methods did not provide conclusive grounds for selecting a preferred number of clusters from the given dataset, and the Akaike information criterion (AIC) and the Bayes Information criterion (BIC), which penalise increases in model likelihoods with the number of parameters, showed continued improvements for numbers of clusters above 50. In effect, the higher the number of clusters, the more granular the observations became, but at the risk of becoming challenging to visualise or too convoluted to interpret. Therefore, the number of clusters has been constrained to 21 for the sake of the visualisations in the proceeding plots. The caveat is that this is not an `optimal' number and that additional details can be recovered by increasing the number of clusters or clustering the data in successive stages to `drill-down' into areas of interest.

\begin{figure}[h]
  \centering
  \includegraphics[width=\textwidth, keepaspectratio]{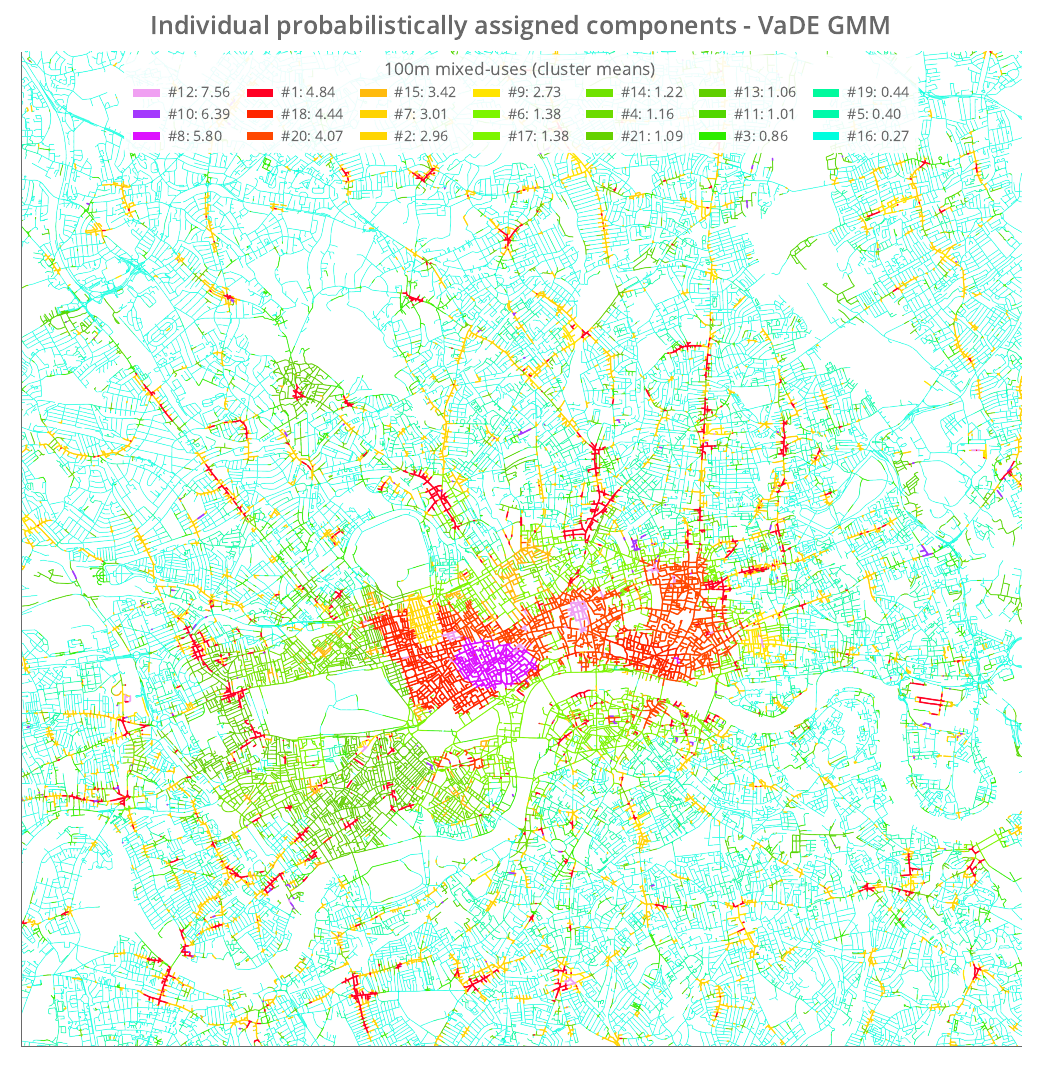}
  \caption[Figure S7: Geographic plot of VaDE clusters.]{Geographic plot of VaDE clusters. Assignment is based on the cluster with the highest probability for a given point, with colours gradated according to the average local mixed-uses for a given cluster.}\label{fig:VaDE_clusters}
 \end{figure}

\subsection{Supplementary Section 4: Variables used for model training.}

\begin{table}[ht]
\centering\footnotesize
\begin{tabular}{ p{1.5cm} | p{5cm} p{2cm} p{2cm} }
  &
  Variables
  &
  Distances
  &
  Data Source
  \\
  \midrule
  Centrality data
  &
  Node harmonic angular closeness centrality\newline
  Node betweenness centrality
  &
  100m to 800m
  &
  OS Open Roads
  \\
  \midrule
  Land-use data
  &
  Mixed-use richness\newline
  Eating establishment accessibility\newline
  Drinking establishment accessibility\newline
  Commercial venue accessibility\newline
  Manufacturing location accessibility\newline
  Food retail shop accessibility\newline
  Other retail shop accessibility\newline
  Transportation or station accessibility\newline
  Total landuse accessibility
  &
  100m to 800m
  &
  OS POI
  \\
  \midrule
  Statistical Data
  &
  Population density\newline
  Dwelling density
  &
  100m to 800m
  &
  ONS Census Data
  \\
\end{tabular}
\caption{Table S1: Summary of source variables used for dimensionality reduction.}\label{table:dim-red-vars}
\end{table}

\begin{table}[ht]
\centering\footnotesize
\begin{tabular}{ p{1.5cm} | p{5cm} p{2cm} p{2cm} }
  &
  Variables
  &
  Distances
  &
  Data Source
  \\
  \midrule
  Centrality data
  &
  Segment density\newline
  Segment beta centrality (gravity index)\newline
  Segment harmonic angular closeness centrality\newline
  Segment betweenness centrality\newline
  Segment angular betweenness centrality
  &
  100m to 800m
  &
  OS Open Roads
  \\
  \midrule
  Land-use data
  &
  Mixed-use richness\newline
  Accommodation establishment accessibility\newline
  Eating establishment accessibility\newline
  Drinking establishment accessibility\newline
  Commercial venue accessibility\newline
  Tourism venue accessibility\newline
  Entertainment venue accessibility\newline
  Government office accessibility\newline
  Manufacturing location accessibility\newline
  Food retail shop accessibility\newline
  Other retail shop accessibility\newline
  Transportation or station accessibility\newline
  Health facility accessibility\newline
  Education institution accessibility\newline
  Park accessibility\newline
  Cultural venue accessibility\newline
  Sport venue accessibility\newline
  Total landuse accessibility
  &
  100m to 800m
  &
  OS POI
  \\
  \midrule
  Statistical Data
  &
  Population density\newline
  Percentage employed\newline
  Dwelling density\newline
  Student population density
  &
  100m to 800m
  &
  ONS Census Data
  \\
\end{tabular}
\caption{Table S2: Summary of source variables used for clustering.}\label{table:clustering-vars}
\end{table}

\clearpage
\section{Citations}
\printbibliography[heading=none]{}

\section{Acknowledgements}
\subsection{PhD}

This paper derives from the author's PhD research at the \emph{Centre for Advanced Spatial Analysis}, \emph{University College London}. The author wishes to acknowledge their PhD supervisors, Prof.~Elsa Arcaute and Prof.~Michael Batty, for their gracious support and feedback throughout the development of this work. The author takes sole responsibility for any oversights or shortcomings contained within this paper.

\subsection{Data}

\begin{flushleft}
The geographical plots and statistical figures in this document have been prepared with use of the following sources of data:\linebreak
\linebreak
\textbf{\emph{Ordnance Survey} \emph{Open Roads}}\linebreak
\emph{Contains OS data © Crown copyright and database right 2021.}\linebreak
\linebreak
\textbf{\emph{Ordnance Survey} \emph{Points of Interest} data}\linebreak
\emph{This material includes data licensed from PointX© Database Right/Copyright 2021.}\linebreak
\emph{Ordnance Survey © Crown Copyright 2021. All rights reserved. Licence number 100034829.}\linebreak
\linebreak
\textbf{\emph{UK Data Service} / \emph{Office for National Statistics} census data}\linebreak
\emph{Contains National Statistics data © Crown copyright and database right 2021.} \linebreak
\emph{Contains OS data © Crown copyright and database right (2021).}\linebreak
\end{flushleft}
\end{document}